\documentclass[aps,prl,floatfix,superscriptaddress,twocolumn,10pt]{revtex4-1}
\usepackage{graphicx}
\usepackage{dcolumn}
\usepackage{bm}
\usepackage{color}
\usepackage{amsmath,amsfonts,amssymb}
\usepackage{graphicx}
\usepackage[english]{babel}
\usepackage{color}
\usepackage{booktabs,longtable}
\usepackage{natbib}
\usepackage{textgreek} 
\usepackage{upgreek} 
\usepackage{mathrsfs}

\usepackage[colorlinks]{hyperref}
\hypersetup{
                pdfstartview={FitH},
                linkcolor=blue,
                citecolor=blue,
                filecolor=blue,
                urlcolor=blue
}

\newcommand\varpm{\mathbin{\vcenter{\hbox{
  \oalign{\hfil$\scriptstyle+$\hfil\cr
          \noalign{\kern-.3ex}
          $\scriptscriptstyle({-})$\cr}
}}}}
\newcommand\varmp{\mathbin{\vcenter{\hbox{
  \oalign{\hfil$\scriptscriptstyle-$\hfil\cr
          \noalign{\kern-.3ex}
          $\scriptstyle({+})$\cr}
}}}}

\begin{document}

\begin{sloppypar}

\title{Coexistence of ferromagnetism and spin-orbit coupling by incorporation of platinum in two-dimensional VSe$_2$}

\author{E. V\'elez-Fort}
\affiliation{Univ. Grenoble Alpes, CEA, CNRS, Grenoble INP, IRIG-SPINTEC, 38000 Grenoble, France}
\author{A. Hallal}
\affiliation{Univ. Grenoble Alpes, CEA, CNRS, Grenoble INP, IRIG-SPINTEC, 38000 Grenoble, France}
\author{R. Sant}
\affiliation{ESRF, The European Synchrotron, 38043 Grenoble, France}
\author{T. Guillet}
\affiliation{Univ. Grenoble Alpes, CEA, CNRS, Grenoble INP, IRIG-SPINTEC, 38000 Grenoble, France}
\author{K. Abdukayumov}
\affiliation{Univ. Grenoble Alpes, CEA, CNRS, Grenoble INP, IRIG-SPINTEC, 38000 Grenoble, France}
\author{A. Marty}
\affiliation{Univ. Grenoble Alpes, CEA, CNRS, Grenoble INP, IRIG-SPINTEC, 38000 Grenoble, France}
\author{C. Vergnaud}
\affiliation{Univ. Grenoble Alpes, CEA, CNRS, Grenoble INP, IRIG-SPINTEC, 38000 Grenoble, France}
\author{J.-F. Jacquot}
\affiliation{Univ. Grenoble Alpes, CEA, IRIG-SYMMES, 38000 Grenoble, France}
\author{D. Jalabert}
\affiliation{Univ. Grenoble Alpes, CEA, IRIG-MEM, 38000 Grenoble, France}
\author{J. Fujii}
\affiliation{Istituto Officina dei Materiali (IOM)-CNR, Laboratorio TASC, Area Science Park, S.S.14, Km 163.5, 34149 Trieste, Italy}
\author{I. Vobornik}
\affiliation{Istituto Officina dei Materiali (IOM)-CNR, Laboratorio TASC, Area Science Park, S.S.14, Km 163.5, 34149 Trieste, Italy}
\author{J. Rault}
\affiliation{Synchrotron SOLEIL, L’Orme des Merisiers, Saint-Aubin, 91192 Gif-sur-Yvette, France}
\author{N. B. Brookes}
\affiliation{ESRF, The European Synchrotron, 38043 Grenoble, France}
\author{D. Longo}
\affiliation{Synchrotron SOLEIL, L’Orme des Merisiers, Saint-Aubin, 91192 Gif-sur-Yvette, France}
\author{P. Ohresser}
\affiliation{Synchrotron SOLEIL, L’Orme des Merisiers, Saint-Aubin, 91192 Gif-sur-Yvette, France}
\author{A. Ouerghi}
\affiliation{Univ. Paris-Saclay, CNRS, Centre de Nanosciences et de Nanotechnologies, 91120, Palaiseau}
\author{J.-Y. Veuillen}
\affiliation{Univ. Grenoble Alpes, CNRS, Grenoble INP, Institut NEEL, 38000 Grenoble, France}
\author{P. Mallet}
\affiliation{Univ. Grenoble Alpes, CNRS, Grenoble INP, Institut NEEL, 38000 Grenoble, France}
\author{H. Boukari}
\affiliation{Univ. Grenoble Alpes, CNRS, Grenoble INP, Institut NEEL, 38000 Grenoble, France}
\author{H. Okuno}
\affiliation{Univ. Grenoble Alpes, CEA, IRIG-MEM, 38000 Grenoble, France}
\author{M. Chshiev}
\affiliation{Univ. Grenoble Alpes, CEA, CNRS, Grenoble INP, IRIG-SPINTEC, 38000 Grenoble, France}
\affiliation{Institut Universitaire de France, 75231, Paris, France}
\author{F. Bonell}
\affiliation{Univ. Grenoble Alpes, CEA, CNRS, Grenoble INP, IRIG-SPINTEC, 38000 Grenoble, France}
\author{M. Jamet}
\affiliation{Univ. Grenoble Alpes, CEA, CNRS, Grenoble INP, IRIG-SPINTEC, 38000 Grenoble, France}
\date{\today}

\begin{abstract}
We report on a novel material, namely two-dimensional (2D) V$_{1-x}$Pt$_x$Se$_2$ alloy, exhibiting simultaneously ferromagnetic order and Rashba spin-orbit coupling. While ferromagnetism is absent in 1T-VSe$_2$ due to the competition with the charge density wave phase, we demonstrate theoretically and experimentally that the substitution of vanadium by platinum in VSe$_2$ (10-50 \%) to form an homogeneous 2D alloy restores ferromagnetic order with Curie temperatures of 6 K for 5 monolayers and 25 K for one monolayer of V$_{0.65}$Pt$_{0.35}$Se$_2$. Moreover, the presence of platinum atoms gives rise to Rashba spin-orbit coupling in (V,Pt)Se$_2$ providing an original platform to study the interplay between ferromagnetism and spin-orbit coupling in the 2D limit.
\end{abstract}

\maketitle

Van der Waals ferromagnets have recently attracted a lot of attention \cite{Gong2019}. From a fundamental point of view, when thinned down to one monolayer (ML), they raised the new paradigm of two-dimensional (2D) ferromagnetism. According to the Mermin-Wagner theorem \cite{Mermin1966}, long-range magnetic order vanishes at finite temperature in 2D Heisenberg spin lattices due to the excitation of long-wavelength spin waves. The introduction of magnetic anisotropy introduced by crystalline symmetries or the application of a magnetic field can restore ferromagnetism at finite temperature in 2D systems  \cite{Gibertini2019}. Van der Waals ferromagnets can be easily stacked with other 2D materials forming well-defined interfaces to study proximity effects \cite{Seyler2018} and conceive new spintronic devices \cite{Wang2018}. Ferromagnetism in 2D materials can be induced by doping with magnetic elements like in Mn-doped MoSe$_2$ \cite{Zhang2015,Gay2021,Dau2019} or V-doped WSe$_2$ \cite{Mallet2020,Pham2020} whereas intrinsic ferromagnetism was first discovered in insulating CrI$_3$ \cite{Huang2017} and semiconducting Cr$_2$Ge$_2$Te$_6$ \cite{Gong2017} in 2017. Today 2D ferromagnets form a vast class of materials including metals such as Fe$_3$GeTe$_2$ \cite{Fei2018,Tan2018,Deng2018}. In this family, transition metal (TM) tellurides exhibit the highest Curie temperatures whereas inducing ferromagnetism in TM selenides still remains a challenge despite few experimental realizations \cite{Nakano2019,Li2021,O'Hara2018}. In this respect, 1T-VSe$_2$ is of particular interest. Although it was initially claimed as a room-temperature 2D ferromagnet \cite{Bonilla2018}, it was demonstrated that the competition with the charge-density-wave (CDW) order prevents ferromagnetism to set in \cite{Feng2018}. As a consequence, a small perturbation of the 1T-VSe$_2$ crystal might drive it either to the ferromagnetic or the CDW phase.\\

In this Letter, we investigate the effect of platinum incorporation in 1T-VSe$_2$ by substitution of vanadium both experimentally and by first principles calculations. 2D V$_{1-x}$Pt$_x$Se$_2$ alloys with $0<x<0.5$ were grown by molecular beam epitaxy (MBE) on epitaxial graphene on SiC(0001). Careful structural analysis shows that we obtained homogeneous 2D alloys with the lattice parameters following the Vegard's law. Magnetic measurements using superconducting quantum interference device (SQUID), x-ray magnetic circular dichroism (XMCD) and anomalous Hall effect (AHE) demonstrate the ferromagnetic order of 1 and 5 ML of V$_{1-x}$Pt$_x$Se$_2$ with $0<x<0.5$ as predicted by first principles calculations. Moreover, spin and angle resolved photoemission spectroscopy (spin-ARPES) on 1 ML of V$_{0.65}$Pt$_{0.35}$Se$_2$ revealed an in-plane spin texture due to the Rashba spin-orbit coupling. This is a consequence of the presence of local dipole within the monolayer as already observed in pure PtSe$_2$ \cite{Yan2017} and on enhanced spin-orbit interaction induced by the presence of Pt. Such coexistence of ferromagnetism and Rashba-like spin-orbit coupling was theoretically predicted in Janus monolayers of magnetic TM dichalcogenides and identified as all-in-one platform for spin-orbit torque \cite{Smaili2020}. Here, we demonstrate experimentally that 2D V$_{1-x}$Pt$_{x}$Se$_2$ constitutes such a model system exhibiting both ferromagnetism and spin texture.

\begin{figure}[t!]
\includegraphics[width=0.48\textwidth]{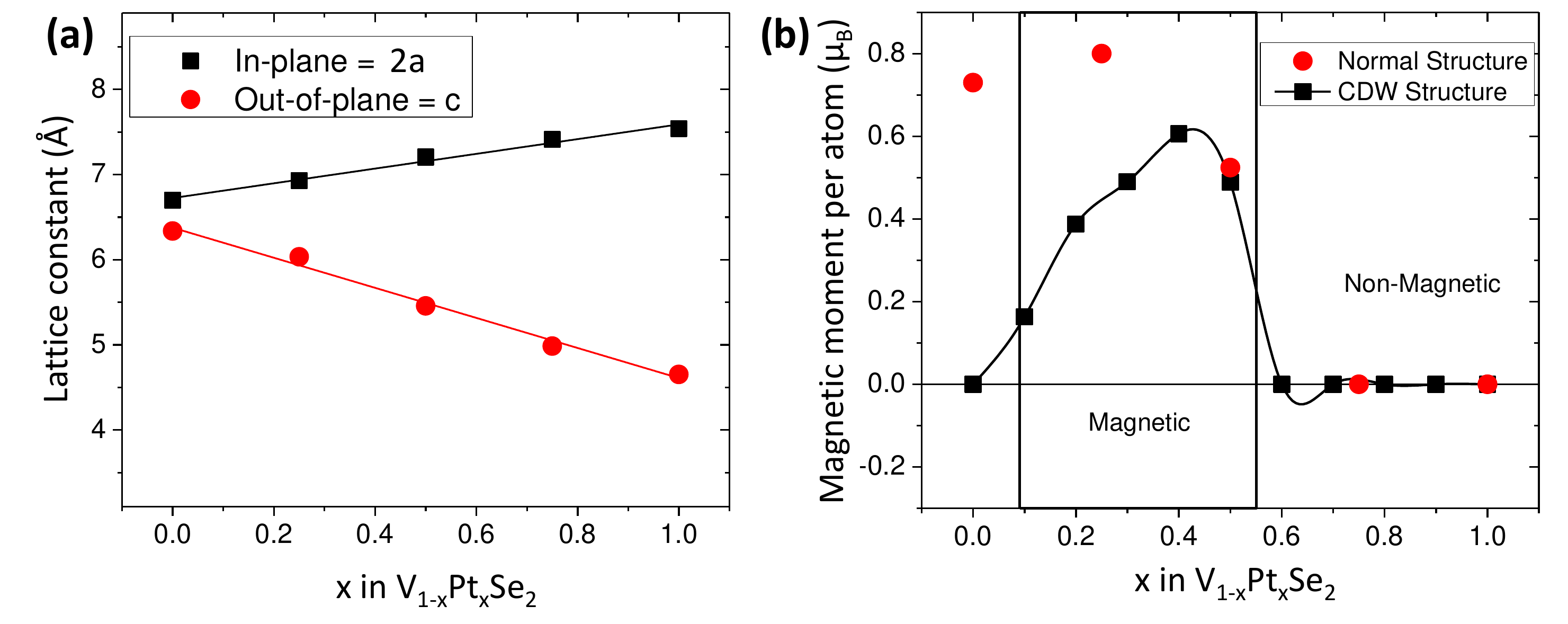} 
\caption{(color online) (a) Calculated in-plane (2a, black squares) and out-of-plane (c, red dots) lattice parameters of 1T-V$_{1-x}$Pt$_x$Se$_2$ as a function of the Pt content $x$. (b) Magnetic moment per metallic atom (V and Pt) in V$_{1-x}$Pt$_x$Se$_2$ as a function of the Pt content $x$ for both normal and CDW structures.}
\label{Fig1}
\end{figure}

First, we present ab initio calculations performed to study the structural and electronic properties of V$_{1-x}$Pt$_x$Se$_2$ alloys using the Vienna Ab-initio Simulation Package (VASP)\cite{Kresse1996} employing projector augmented wave (PAW)\cite{Blochl1994} pseudopotentials. The calculations details are given in Refs. \cite{Sup,Ceperley1980,Coelho2019}. In Fig.~\ref{Fig1}a, the calculated in-plane and out-of-plane lattice parameters follow the Vegard's law as expected for a solid solution made of an homogeneous mixture of 1T-VSe$_2$ and 1T-PtSe$_2$. The magnetic properties of both normal and CDW structures are reported in Fig.~\ref{Fig1}b. One can see that the normal structure possesses itinerant ferromagnetism for up to $x=0.5$ with a maximum magnetic moment per metallic atom (V and Pt) of 0.8 $\mu_B$ where $\mu_B$ is the Bohr magneton. Moreover the spin polarization at the Fermi level increases with the Pt content reaching 96 \% for V$_{0.5}$Pt$_{0.5}$Se$_2$ \cite{Sup}. When the CDW structure is considered, the ferromagnetic order disappears for $x=0$ as already demonstrated experimentally \cite{Feng2018} and theoretically \cite{Coelho2019}. However, when increasing the Pt content, $x$, ferromagnetism appears again for $0.1<x<0.5$ with a maximum magnetic moment of 0.6 $\mu_B$ per metallic atom (V and Pt).\\

Following these predictions, we used MBE to grow V$_{1-x}$Pt$_x$Se$_2$ on epitaxial graphene on SiC with $0<x<0.5$. V and Pt were evaporated from e-gun evaporators and selenium from an effusion cell. The base pressure in the MBE reactor was 5$\times$10$^{-10}$ mbar. The Gr/SiC substrate was first outgassed under ultrahigh vacuum at 800$^{\circ}$C during 30 minutes and maintained at 280$^{\circ}$C during the growth. The V and Pt deposition rates were adjusted using quartz balance monitors to reach the targeted composition. The Se pressure was fixed at 1.0$\times$10$^{-6}$ mbar in excess with respect to the V and Pt deposition rates by a factor $\approx$15. After the growth, the layers were annealed at 500$^{\circ}$C during 20 minutes and capped at room temperature with amorphous selenium for \textit{ex situ} characterizations to protect the layers from oxidation during transfer in air. The surface crystalline order and morphology were monitored \textit{in situ} by reflection high energy electron diffraction (RHEED).

\begin{figure}[t!]
\includegraphics[width=0.48\textwidth]{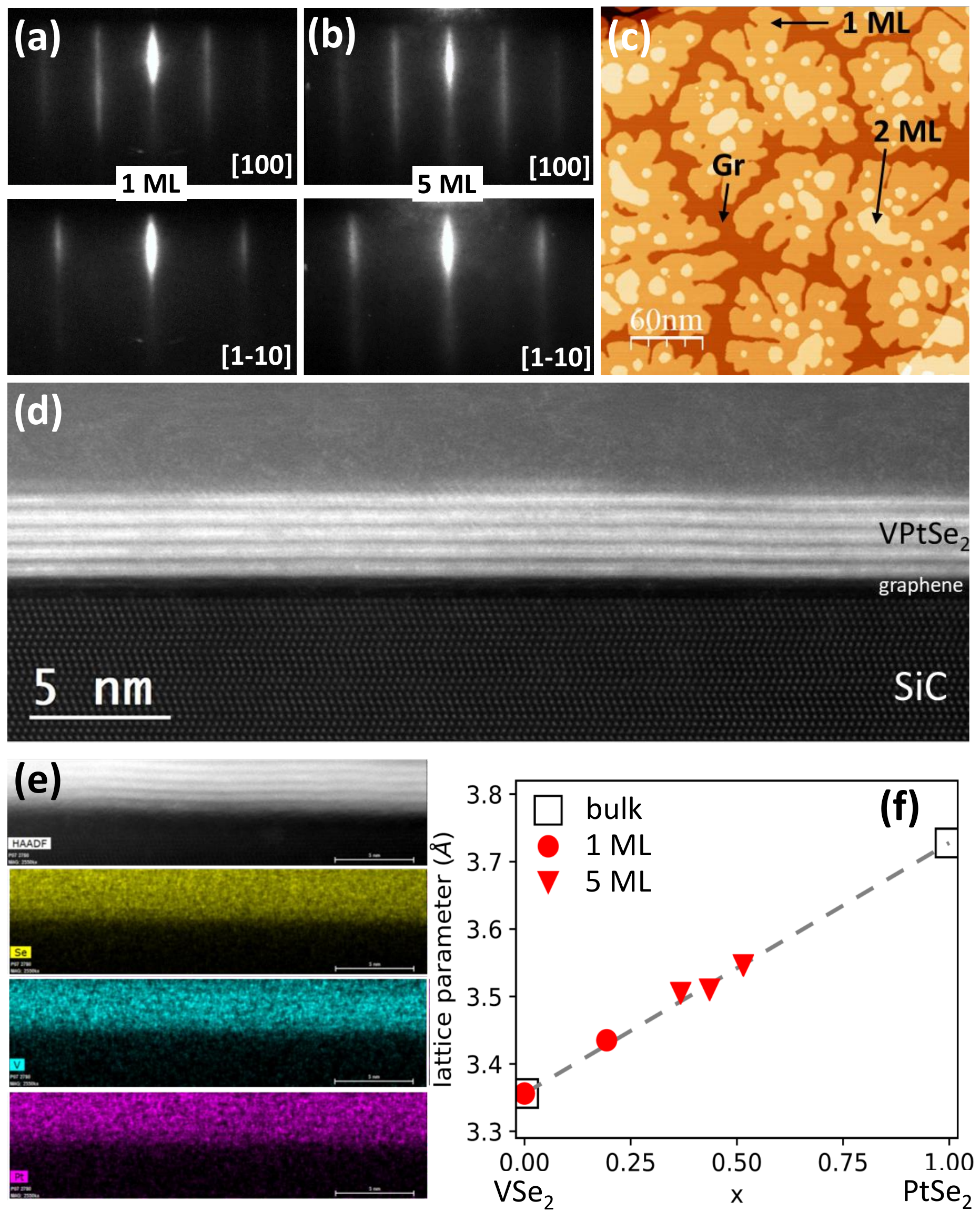} 
\caption{(color online) (a) and (b) RHEED patterns for  1 ML and 5 ML of V$_{0.65}$Pt$_{0.35}$Se$_2$ along two different azimuths [100] and [1-10] separated by 30$^{\circ}$. (c) 300$\times$300 nm STM image recorded at 8.5 K of 1 ML of V$_{0.65}$Pt$_{0.35}$Se$_2$. The sample bias was -2.5 V and the tunneling current 100 pA. (d) HAADF-STEM images in cross-section of 5 ML of V$_{0.65}$Pt$_{0.35}$Se$_2$. (e) Energy dispersive x-ray (EDX) chemical analysis of the HAADF image for the selenium (yellow), vanadium (blue) and platinum (purple) elements. (f) In-plane lattice parameter of V$_{1-x}$Pt$_x$Se$_2$ as a function of the Pt content $x$ for bulk references (open squares), 1 ML (red dots) and 5 ML (red triangles). The Pt content was calibrated thanks to Rutherford Backscattering measurements and the lattice parameters deduced from in-plane x-ray diffraction.}
\label{Fig2}
\end{figure}

In Fig.~\ref{Fig2}a and ~\ref{Fig2}b, RHEED patterns for  1 ML and 5 ML of V$_{0.65}$Pt$_{0.35}$Se$_2$ are shown along two different azimuths separated by 30$^{\circ}$. The number of deposited layers was controlled \textit{in situ} by recording the RHEED intensity oscillations during the growth. We first notice that both 1 ML and 5 ML of V$_{0.65}$Pt$_{0.35}$Se$_2$ grow epitaxially on graphene as confirmed by x-ray diffraction \cite{Sup}. The epitaxial relationship is (V,Pt)Se$_2$(110)$\vert\vert$Gr(110)$\vert\vert$SiC(100). The morphology of 1 ML is shown in Fig.~\ref{Fig2}c as measured by low temperature scanning tunneling microscopy (STM). For this observation, the sample was transferred in air and the Se capping layer thermally removed in the microscope prior to the measurements. This image and height profiles (not shown) demonstrate the monolayer character of the film with 1 ML of V$_{0.65}$Pt$_{0.35}$Se$_2$ partially covered by a second layer. In Fig.~\ref{Fig2}d, we used scanning transmission electron microscopy (STEM) in the high-angle annular dark field mode (HAADF) to image the 5 ML film in cross-section with atomic resolution. In this image, we clearly distinguish the SiC substrate, the graphene layer and the 5 ML of V$_{0.65}$Pt$_{0.35}$Se$_2$ with sharp interfaces. From higher magnification images, we could not detect the presence of atoms in the vdW gaps, both the ones between (V,Pt)Se$_2$ layers and the one at the interface with graphene, confirming the vdW layered character of V$_{0.65}$Pt$_{0.35}$Se$_2$ without atomic intercalation. In Fig.~\ref{Fig2}e, the chemical analysis corresponding to the cross-section in Fig.~\ref{Fig2}d shows that V, Pt and Se atoms are homogeneously distributed in the V$_{0.65}$Pt$_{0.35}$Se$_2$ layers. For both 1 ML and 5 ML films, we could not detect any secondary phase, clusters or inhomogeneities neither by STEM, STM nor by x-ray diffraction confirming that we succeeded in growing an homogeneous vanadium-platinum diselenide alloy. Finally, all the in-plane lattice parameters measured by x-ray diffraction for 1 ML, 5 ML and bulk V$_{1-x}$Pt$_x$Se$_2$ as a function of the Pt content $x$ are reported in Fig.~\ref{Fig2}f. All the data fall on a single line demonstrating that the in-plane lattice parameter of V$_{1-x}$Pt$_x$Se$_2$ follows the Vegard's law: $a$(V$_{1-x}$Pt$_x$Se$_2$)=$(1-x)a$(VSe$_2$)$+xa$(PtSe$_2$), in good agreement with first principles calculations. We further extracted the out-of-plane lattice parameter $c$ from $\theta$/2$\theta$ x-ray diffraction of the 5 ML film and found $c=0.59$ nm which is only 3.3 \% larger than the expected value from the Vegard's law and theoretical predictions \cite{Sup}.

\begin{figure*}[t!]
\includegraphics[width=\textwidth]{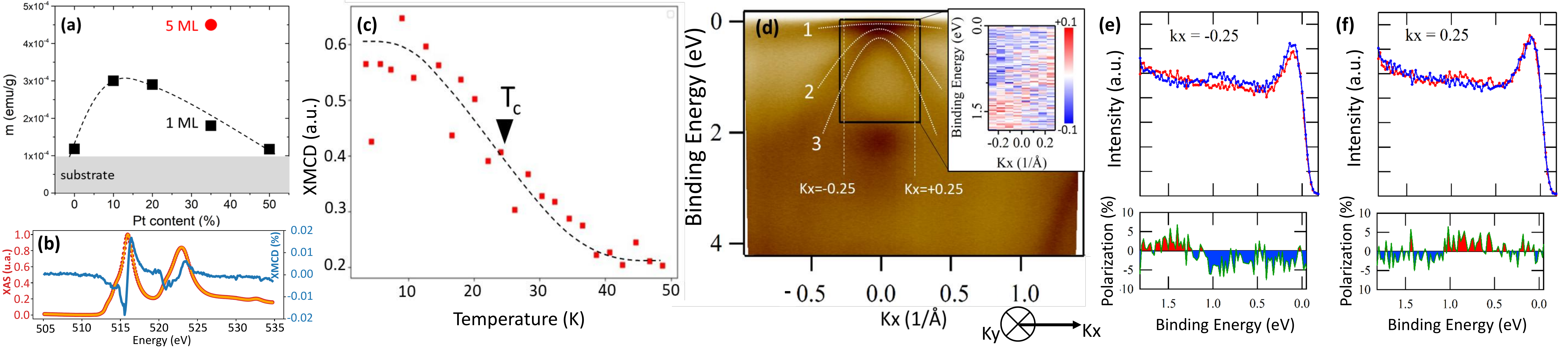} 
\caption{(color online) (a) Ferromagnetic component extracted from hysteresis loops recorded at 5 K by SQUID as a function of the Pt content in \% \cite{Sup}. The measured magnetization has been normalized to the substrate mass. The field is applied out-of-plane. Black squares are for monolayers and the red dot for 5 ML. The dashed black line is a guide for the eye. Several Gr/SiC substrates were measured and the corresponding ferromagnetic signals fall within the grey area. (b) XAS (red: I$_{\sigma_+}$(+9T)+I$_{\sigma_-}$(-9T), orange: I$_{\sigma_+}$(-9T)+I$_{\sigma_-}$(+9T)) and resulting XMCD (blue) spectra for 1 ML of V$_{0.65}$Pt$_{0.35}$Se$_2$ recorded at 7 K in total electron yield mode. The field is applied perpendicular to the film and XMCD was recorded in normal incidence. The XAS spectra are normalized to their maximum value. (c) XMCD signal as a function of temperature. The XMCD signal was determined by integration in the range 510-520 eV at each temperature under 9 T. The black dashed line is a guide for the eye. (d) Electronic structure of 1 ML V$_{0.65}$Pt$_{0.35}$Se$_2$ measured by spin-ARPES (T=77 K) with vertical linear polarization ($h\nu$=70 eV). The map displays the angle-resolved valence band spectra along the $\Gamma$-K direction. Inset: spin polarization map (red: spin in-plane along +K$_y$, blue: spin in-plane along -K$_y$). The corresponding area is indicated by a black rectangle in the ARPES map. (e) and (f) Spin-resolved energy distribution curves taken at $k_x$=-0.25 1/\AA\ and $k_x$=+0.25 1/\AA\ respectively along with the corresponding spin polarization.}

\label{Fig3}
\end{figure*}

In Fig.~\ref{Fig3}a, we show the results of SQUID measurements where we recorded hysteresis loops at 5 K for different Pt contents (the raw data can be found in   Ref. \cite{Sup}). To avoid any degradation of the films, we kept the amorphous Se capping layer. For each composition, we first removed the diamagnetic contribution from the substrate measured at 100 K and then subtracted a linear contribution of positive slope (estimated at high field) at 5 K corresponding to a paramagnetic component coming from impurities in the substrate and any magnetic contribution from the layer. The remaining magnetic signal is saturating and corresponds to the ferromagnetic component reported in Fig.~\ref{Fig3}a. We clearly see that a ferromagnetic signal above the substrate background emerges between 0 and 50 \% of platinum as expected from first principles calculations in Fig.~\ref{Fig1}b. Moreover, the ferromagnetic signal is 4.4 times larger in  5 ML than in 1 ML of V$_{0.65}$Pt$_{0.35}$Se$_2$ after subtracting the substrate background signal. This value is close to 5, expected using a simple scaling law. Considering the magnetic moment and volume of the 5 ML sample as well as the lattice parameter for the V$_{0.65}$Pt$_{0.35}$Se$_2$ composition, we find a magnetic moment of $\approx$0.4 $\mu_B$ per metal atom. This value is comparable to the calculated one reported in Fig.~\ref{Fig1}b. X-ray absorption (XAS) and XMCD measurements at 7 K performed at the European Synchrotron Radiation Facility (ESRF, beamline ID32 \cite{Kummer2016}) and at SOLEIL (beamline DEIMOS \cite{Ohresser2014}) are shown in Fig.~\ref{Fig3}b for 1 ML of V$_{0.65}$Pt$_{0.35}$Se$_2$. The applied magnetic field was 9 T. The XMCD signal at the L$_3$ edge is of the order of 1.9 \% proving that 1 ML of V$_{0.65}$Pt$_{0.35}$Se$_2$ is magnetic despite its two-dimensional character. Unfortunately, we cannot compare the magnetic moment measured by XMCD with the theoretical value of Fig.~\ref{Fig1}b since: (i) the sum rules are not applicable to vanadium due to its very weak spin-orbit coupling and (ii) the magnetic moment of Pt atoms could not be measured because the energy range was not accessible. To estimate the Curie temperature, we recorded the temperature dependence of the XMCD signal as shown in Fig.~\ref{Fig3}c. Starting from 15 K the signal slowly decreases until it reaches a plateau at 40 K. We attribute this to a magnetic transition occurring within this range of temperature. The  inflection point close to 25 K can be attributed to the Curie temperature (T$_{\textup{C}}$) of the layer.


We also probed the electronic structure of (V,Pt)Se$_2$ by ARPES at the Cassioppee beamline of SOLEIL and spin-ARPES performed at the APE-NFFA beamline \cite{Panaccione2009} at the Elettra synchrotron facility by use of a Scienta DA30 hemispherical analyzer combined with a VLEED type spin detector \cite{Bigi2017}. For these measurements, the Se capping layer was thermally removed. The ARPES spectra are in good agreement with the band structure derived from first principles calculations \cite{Sup}. In Fig.~\ref{Fig3}d, we present the measured valence band structure and spin texture of 1 ML of V$_{0.65}$Pt$_{0.35}$Se$_2$ along the $\Gamma$-K direction of the first Brillouin zone. Close to the Fermi level E$_{\textup{F}}$, we can distinguish 3 different bands, 1-3, highlighted by white dashed lines. Band 1 is nearly flat and in close proximity to the Fermi level which indicates that these states are involved in magnetotransport properties. Bands 2 and 3 exhibit a hole-like character and a parabolic dispersion. The measured spin texture of these bands is shown in the inset corresponding to the area in the ARPES map delimited by the black rectangle. The spin component is in-plane and perpendicular to $k_x$ and there is evident inversion of the sign of the spin polarization across $k_x=0$, as typically observed for the case of spin-momentum locking. Spin-resolved energy distribution curves taken at $k_x$=-0.25 1/\AA\ and $k_x$=+0.25 1/\AA\ (Fig.~\ref{Fig3}e and ~\ref{Fig3}f) show that the sign inversion of the spin polarization in the map is due to the inversion of the spin polarization of two broad peaks. The first (resp. second) one in the 0.6-1.2 eV (resp. 1.2-1.8 eV) binding energy range corresponds to the hole-like band noted 2 (resp. 3) in Fig.~\ref{Fig3}d. The spin polarization reaches a maximum value of 5 \%. This observation suggests that both bands exhibit a helical in-plane spin texture of opposite chirality. The same spin textures attributed to spin-layer locking by local Rashba effect were observed in pure PtSe$_2$ monolayer with much stronger spin polarization \cite{Yan2017}. We conclude that the same effect takes place in the monolayer of V$_{0.65}$Pt$_{0.35}$Se$_2$ by the incorporation of heavy Pt in VSe$_2$. This 2D material thus exhibits both ferromagnetism and in-plane spin texture where exchange interaction and Rashba spin-orbit coupling coexist. 

\begin{figure}[t!]
\includegraphics[width=0.48\textwidth]{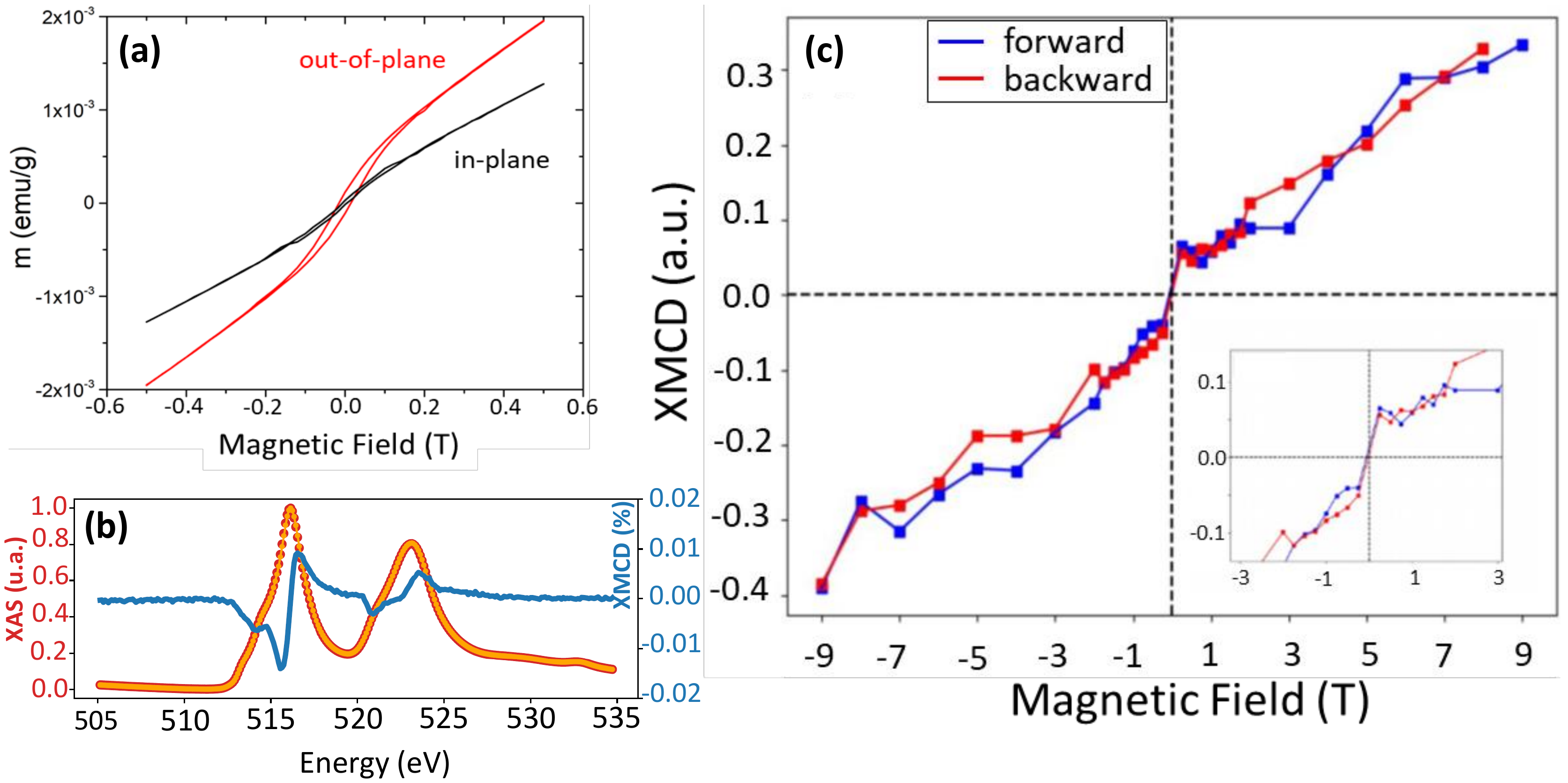} 
\caption{(color online) (a) Hysteresis loops recorded at 5 K by SQUID with the field applied in-plane (black) and out-of-plane (red) for 5 ML of V$_{0.65}$Pt$_{0.35}$Se$_2$. (b) XAS (red: I$_{\sigma_+}$(+9T)+I$_{\sigma_-}$(-9T), orange: I$_{\sigma_+}$(-9T)+I$_{\sigma_-}$(+9T)) and resulting XMCD (blue) spectra for 5 ML of V$_{0.65}$Pt$_{0.35}$Se$_2$ recorded at 7 K in total electron yield. The field is applied perpendicular to the film and XMCD was recorded in normal incidence. The XAS spectra are normalized to their maximum value. (c) Magnetization curve for 5 ML of V$_{0.65}$Pt$_{0.35}$Se$_2$ recorded at 7 K for increasing field (forward, blue) and decreasing field (backward, red). Inset: zoom in between -3 T and +3 T. The XMCD signal was determined by integration in the range 510-520 eV at each magnetic field value.}

\label{Fig4}
\end{figure}

The hysteresis loops at 5 K for 5 ML of V$_{0.65}$Pt$_{0.35}$Se$_2$ with the field applied in-plane and out-of-plane are reported in Fig.~\ref{Fig4}a. Only the diamagnetic signal from the substrate at 100 K was removed. Two contributions can be well identified: one ferromagnetic component inferred from the hysteresis behavior with clear out-of-plane anisotropy and one linear component inferred from the linear field dependence that can arise from paramagnetic impurities in the substrate and any magnetic contribution from the layer. The same out-of-plane anisotropy is visible in SQUID raw data for compositions ranging from 20 to 50 \% of Pt \cite{Sup}. We also perfomed XMCD measurements and Fig.~\ref{Fig4}b shows the XAS and XMCD spectra at the L$_{2,3}$ edges of vanadium recorded at 7 K under a magnetic field of 9 T (additional XMCD data on 5 ML of V$_{0.5}$Pt$_{0.5}$Se$_2$ can be found in Ref. \cite{Sup}). The XMCD signal (1.2 \% at the L$_3$ edge) is lower than for 1 ML but clearly demonstrates that vanadium atoms still carry a magnetic moment. The magnetization curve deduced from XMCD measurements is reported in Fig.~\ref{Fig4}c. We observe a weak ferromagnetic component corresponding to the step near zero field and a linear contribution in agreement with SQUID measurements. It demonstrates that part of the SQUID linear magnetic signal originates from the V$_{0.65}$Pt$_{0.35}$Se$_2$ layer. Such coexistence of a weak ferromagnetic signal and a non-saturating magnetic contribution indicates that 7 K is close to the Curie temperature of the film.

\begin{figure}[t!]
\includegraphics[width=0.48\textwidth]{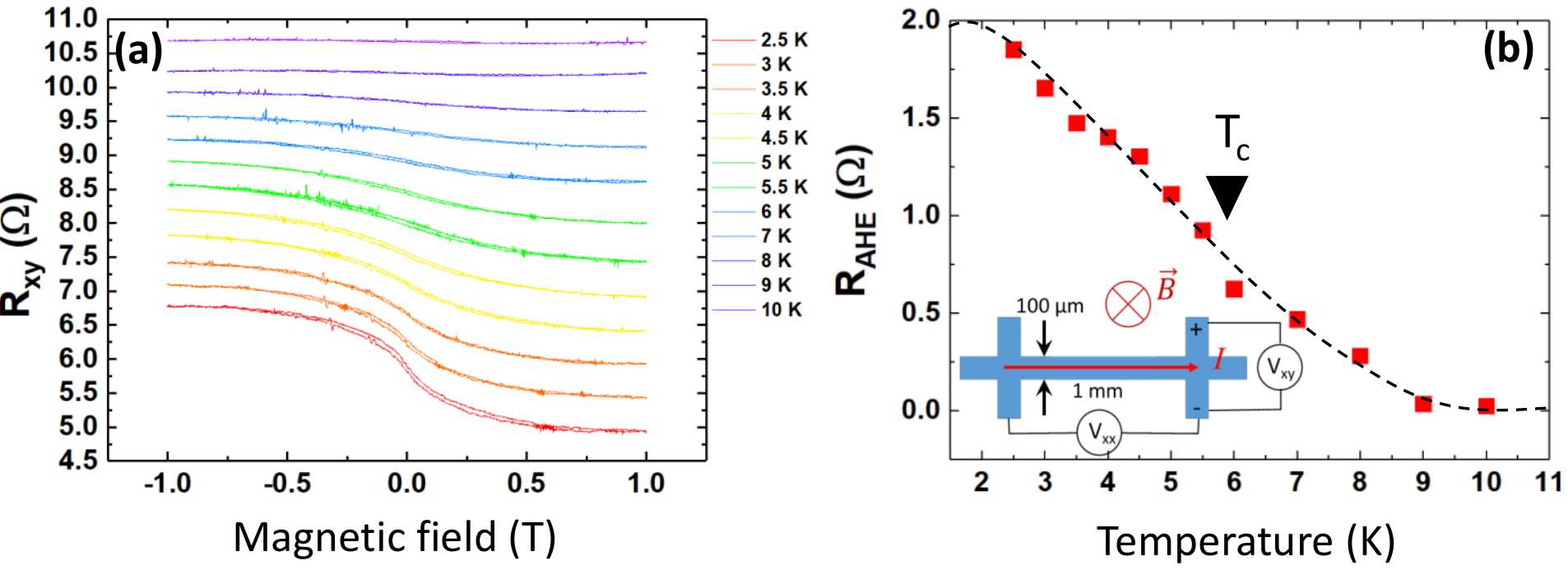} 
\caption{(color online) (a) Transverse resistance R$_{\textup{xy}}$=V$_{\textup{xy}}$/I as a function of the applied magnetic field for different temperatures (the longitudinal resistance R$_{\textup{xx}}$ is reported in Ref. \cite{Sup}). The curves are shifted vertically for clarity. (b) R$_{\textup{AHE}}$=R$_{\textup{xy}}^{\textup{max}}$=R$_{\textup{xy}}$(1T) as a function of temperature. The dashed black line is a guide for the eye. Inset: sketch of the Hall bar.}
\label{Fig5}
\end{figure}

In order to estimate this transition temperature, we performed magnetotransport measurements \cite{note}. For this purpose, we grew 5 ML of V$_{0.65}$Pt$_{0.35}$Se$_2$ epitaxially on mica given that electrical measurements of (V,Pt)Se$_2$ were impossible when grown on graphene since most of the current was shunted by the graphene layer. In this respect, mica is insulating while keeping the single crystalline and vdW characters. The details about the growth and Hall bar processing are given in Ref. \cite{Sup}. By applying a perpendicular magnetic field $\vec{B}$ and an electrical current I=10 $\mu$A in the Hall bar sketched in the inset of Fig.~\ref{Fig5}b, we recorded the transverse resistance R$_{\textup{xy}}$ for different temperatures in Fig.~\ref{Fig5}a. Hysteresis loops can be observed for the lowest temperatures and we plot the AHE as a function of temperature in Fig.~\ref{Fig5}b. The signal decreases continuously and vanishes at 9 K. The position of the inflection point gives an estimation of the Curie temperature T$_{\textup{C}}\approx$6 K confirming that SQUID and XMCD measurements in Fig.~\ref{Fig4}a and ~\ref{Fig4}c were performed close to the transition temperature. It should be noticed that, contrary to other vdW ferromagnets like CrI$_3$ \cite{Huang2017}, the magnetic moment per vanadium atom and the Curie temperature are larger in 1 ML than in 5 ML. Though it is out of the scope of this paper, this observation should be clarified in a future work.

In summary, we have grown epitaxially two-dimensional V$_{1-x}$Pt$_x$Se$_2$ alloys with $0<x<0.5$. Careful structural analysis demonstrates that the films are homogeneous with in-plane and out-of-plane lattice parameters following the Vegard's law. The introduction of Pt in VSe$_2$ drives the system into the ferromagnetic phase in agreement with first principles calculations. Ferromagnetism coexists with an helical spin texture in the monolayer form making it an original platform for the study of magnetization manipulation by an electrical current. We hope this paper will trigger more research on such multifunctional 2D materials. This is of paramount importance for the development of new spintronic devices based on this class of materials.\\



The authors acknowledge the financial support from the ANR project MAGICVALLEY (ANR-18-CE24-0007) and the EU H2020 Graphene Flagship. The LANEF Framework (ANR-10-LABX-51-10) is acknowledged for its support with mutualized infrastructure. The authors are grateful to the ESRF and SOLEIL staff for smoothly running the facility. This work has been partly performed in the framework of the nanoscience foundry and fine analysis (NFFA-MIUR Italy Progetti Internazionali) facility.

\end{sloppypar}


\begin{thebibliography}{}
\expandafter\ifx\csname natexlab\endcsname\relax\def\natexlab#1{#1}\fi
\expandafter\ifx\csname bibnamefont\endcsname\relax
  \def\bibnamefont#1{#1}\fi
\expandafter\ifx\csname bibfnamefont\endcsname\relax
  \def\bibfnamefont#1{#1}\fi
\expandafter\ifx\csname citenamefont\endcsname\relax
  \def\citenamefont#1{#1}\fi
\expandafter\ifx\csname url\endcsname\relax
  \def\url#1{\texttt{#1}}\fi
\expandafter\ifx\csname urlprefix\endcsname\relax\def\urlprefix{URL }\fi
\providecommand{\bibinfo}[2]{#2}
\providecommand{\eprint}[2][]{\url{#2}}

\bibitem{Gong2019}
C. Gong and X. Zhang, Science \textbf{363}, eaav4450 (2019).

\bibitem{Mermin1966}
N. D. Mermin and H.  Wagner, Phys. Rev. Lett. \textbf{17}, 1133 (1966).

\bibitem{Gibertini2019}
M. Gibertini, M. Koperski, A. F. Morpurgo and K. S. Novoselov, Nat. Nanotechnol. \textbf{14}, 408 (2019).

\bibitem{Seyler2018}
K. L. Seyler et al., Nano Lett. \textbf{18}, 3823 (2018).

\bibitem{Wang2018}
Z. Wang et al., Nano Lett. \textbf{18}, 4303 (2018).

\bibitem{Zhang2015}
K. Zhang et al., Nano Lett. \textbf{15}, 6586 (2015).

\bibitem{Gay2021}
M. Gay et al., to appear in CRAS 2021.

\bibitem{Dau2019}
M.-T. Dau et al., APL Mater. \textbf{7}, 051111 (2019).

\bibitem{Mallet2020}
P. Mallet, F. Chiapello, H. Okuno, H. Boukari, M. Jamet, and J.-Y. Veuillen, Phys. Rev. Lett. \textbf{125}, 036802 (2020).

\bibitem{Pham2020}
Y. T. H. Pham et al., Adv. Mater. \textbf{32}, 2003607 (2020).

\bibitem{Gong2017}
C. Gong et al., Nature \textbf{546}, 265 (2017).

\bibitem{Huang2017}
B. Huang et al., Nature \textbf{546}, 270 (2017).

\bibitem{Fei2018}
Z. Fei et al., Nat. Mater. \textbf{17}, 778 (2018).

\bibitem{Tan2018}
C. Tan et al., Nat. Commun. \textbf{9}, 1554 (2018).

\bibitem{Deng2018}
Y. Deng et al., Nature \textbf{563}, 94 (2018).

\bibitem{Nakano2019}
M. Nakano et al., Nano Lett. \textbf{19}, 8806 (2019).

\bibitem{Li2021}
B. Li et al., Nat. Mater. (2021). https://doi.org/10.1038/s41563-021-00927-2.

\bibitem{O'Hara2018}
D. J. O'Hara et al., Nano Lett. \textbf{18}, 3125 (2018).

\bibitem{Bonilla2018}
M. Bonilla et al., Nat. Mater. \textbf{13}, 289 (2018).

\bibitem{Feng2018}
J. Feng et al., Nano Lett. \textbf{18}, 4493 (2018).

\bibitem{Yan2017}
M. Yan et al., 2D Mater. \textbf{4}, 045015 (2017).

\bibitem{Smaili2020}
I. Smaili et al., arXiv:2007.07579v1.

\bibitem{Kresse1996}
G. Kresse and J. Furthm\"uller, J. Comput. Mater. Sci. \textbf{6}, 15 (1996).

\bibitem{Blochl1994}
P. E. Bl\"ochl, Phys. Rev. B: Condens. Matter Mater. Phys. \textbf{50}, 17953 (1994).

\bibitem{Ceperley1980}
D. M. Ceperley, B. J. Alder, Phys. Rev. Lett. \textbf{45}, 566 (1980).

\bibitem{Coelho2019}
P. M. Coelho et al., J. Phys. Chem. C \textbf{123}, 14089 (2019).

\bibitem{Sup}
Supplementary information are available at:

\bibitem{Kummer2016}
K. Kummer et al., J. Synchrotron Rad. \textbf{23}, 464 (2016).

\bibitem{Ohresser2014}
P. Ohresser et al., Rev. Sci. Instrum. \textbf{85}, 013106 (2014).

\bibitem{Panaccione2009}
G. Panaccione et al., Rev. Sci. Instrum. \textbf{80}, 43105 (2009).

\bibitem{Bigi2017}
C Bigi et al., J. Synchrotron Rad. \textbf{24}, 750 (2017).

\bibitem{note}
Note that the resistance of 1 ML of V$_{0.65}$Pt$_{0.35}$Se$_2$ was too high to perform electrical measurements, most probably due to the discontinuity of the single layer film.

\end{thebibliography}
\end{document}